\documentclass[prd,twocolumn,showpacs,floatfix,amsmath,nofootinbib,amssymb,floatfix]{revtex4}
\usepackage{graphicx,color,dcolumn,booktabs,bm,multirow}
\usepackage{longtable,lscape}
\usepackage{txfonts}
 \usepackage{enumitem}
\usepackage{overpic}
\usepackage{amssymb}
\usepackage{indentfirst}
\usepackage{feynmf}   %{feynmp}
\usepackage{slashed}  %for Feynman symbols
\usepackage{cases}
\usepackage{color}
\usepackage{multirow}
\usepackage{epstopdf}
\usepackage{graphicx,color,dcolumn,booktabs,bm}

\usepackage[colorlinks,
            citecolor=blue,
            anchorcolor=red,
            menucolor=red,
            linkcolor=red,
            filecolor=red,
            runcolor=red,
            urlcolor=blue,
            frenchlinks=red]{hyperref}
\usepackage{amsmath}
\usepackage{epsfig}
\usepackage{graphicx}
\usepackage{txfonts}

\newcommand{\tccs}{T_{cc\bar{s}}}
\newcommand{\ssqq}{ss\bar{q}\bar{q}}

\newcommand{\ssnn}{ss\bar{n}\bar{n}}
\newcommand{\br}{\boldsymbol{r}}

\begin{document}

\title{Exploring the spectroscopic features of double-strangeness tetraquark states}

\author{Xuejie Liu$^1$}\email[E-mail: ]{1830592517@qq.com}
%\author{Haoming Zheng$^1$}
\author{Haoming Zheng$^1$}\email[E-mail: ]{2402183023@stu.htu.edu.cn}
\author{Dianyong Chen$^{2,4}$\footnote{Corresponding author}}\email[E-mail:]{chendy@seu.edu.cn}
\author{Hongxia Huang$^3$}\email[E-mail:]{hxhuang@njnu.edu.cn}
\author{Jialun Ping$^3$}\email[E-mail: ]{jlping@njnu.edu.cn}
\affiliation{$^1$School of Physics, Henan Normal University, Xinxiang 453007, P. R. China}
\affiliation{$^2$School of Physics, Southeast University, Nanjing 210094, P. R. China}
\affiliation{$^3$Department of Physics, Nanjing Normal University, Nanjing 210023, P.R. China}
\affiliation{$^4$Lanzhou Center for Theoretical Physics, Lanzhou University, Lanzhou 730000, P. R. China}

\date{\today}

\begin{abstract}
 Since the discovery of the $T_{cc}$ double-charm tetaquark by the LHCb collaboration,  the field of the theoretical research on heavy quarks has advanced rapidly, with increasing interest in exploring the light quark sector. In this study, the quark model is employed to systematically analyze the double-strange tetraquark system. Both the meson-meson configuration and diquark-antidiquark configuration are considered. The interactions between hadron pairs under various quantum numbers, as well as the possibilities of bound states and resonances, are evaluated. The results indicate the presence of two bound states, $ \bar{K}^{\ast }\bar{K}$ and $\bar{K}^{\ast }\bar{K}^{\ast}$, with quantum number $I(J^{P})=0(1^{+})$ in single-channel estimations. Additionally, by considering the channel coupling between the two configurations, a bound state with quantum numbers  $I(J^{P})=0(1^{+})$ and a mass of approximately $1310$ MeV is obtained. Moreover,  through the application of Resonance Ground Method, a resonance state is identified in the $I(J^{P})=0(1^{+})$ $\ssqq$ system, with an estimated mass of around 1783 MeV and a decay width of approximately 17 MeV.

\end{abstract}

\pacs{13.75.Cs, 12.39.Pn, 12.39.Jh}
\maketitle

\setcounter{totalnumber}{5}

\section{\label{sec:introduction}Introduction}
Quantum chromodynamics (QCD)  is the foundational theory describing the strong interactions between quarks and gluons.  In the traditional quark model, hadrons consist of mesons, which are composed of a quark-antiquark pair, and baryons, which are composed of three quarks~\cite{Gell-Mann:1964ewy,Lichtenberg:1980rv,ParticleDataGroup:2022pth}. However, according to QCD theory, there exist hadronic states that go beyond the traditonal quark model, such as multi-quark states, hadronic molecules, glueballs, and hybrid states~\cite{Meyer:2015eta,Chen:2016qju,Clement:2016vnl,Guo:2017jvc,Liu:2019zoy,Brambilla:2019esw,Chen:2022asf,Liu:2024uxn}

Since 2003, the experimental observation of several exotic hadronic states that lie beyond the traditional quark model has significantly revitalized interest in the field of exotic hadron research. These states not only expand the diversity of the hadron spectrum but also provide a valuable opportunity to probe the non-perturbative dynamics of the strong interaction through detailed analyses of their properties and structures. To date, the majority of observed exotic states exhibit heavy flavor quantum numbers. For instance, many charmonium-like states, referred to as XYZ states, possess hidden charm components~\cite{Brambilla:2019esw}. Specific examples such as X(3872)~\cite{Belle:2003nnu,BaBar:2004oro,CDF:2003cab,CDF:2009nxk,D0:2004zmu,LHCb:2011zzp,CMS:2013fpt} and $Z_{c}(3900)$~\cite{BESIII:2013ouc,Belle:2013yex} are situated near two-hadron thresholds, yet their precise internal structures remain under debate~\cite{Wang:2021aql,Liu:2013waa,Hosaka:2016pey,Richard:2016eis,Lebed:2016hpi,Olsen:2017bmm,Meng:2022ozq,Chen:2022asf}. A significant recent finding is the identification by the LHCb Collaboration of the double-charm $T_{cc}^{+}(3875)$ state in the  $D^{0}D^{0}\pi^{+}$  invariant mass spectrum~\cite{LHCb:2021vvq}.  Further analysis of  $T_{cc}^{+}(3875)$ state reveals its quark content of $cc\bar{u}\bar{d}$,  identifying it as a doubly charmed tetraqaurk state. Unlike hidden-charm tetraquark states (such as X(3872)),  $T_{cc}^{+}(3875)$ state  does not contain a $c\bar{c}$ pair, and thus quark-antiquark annihilation effects are absent. This observation provides strong evidence supporting the existence of double-charm tetraquark states. Theoretically, there are multiple interpretation of $T_{cc}^{+}(3875)$ state, including it being the compact tetraquark~\cite{Zhang:2021yul,Weng:2021hje,Guo:2021yws,Liu:2023vrk,Meng:2023for,Noh:2023zoq,Mutuk:2023oyz,Wang:2024vjc,Park:2024cic,Li:2023wug}, the hadronic molecule~\cite{Meng:2023jqk,Ma:2023int,Asanuma:2023atv,Sakai:2023syt,Qiu:2023uno,Zhao:2021cvg,Ke:2021rxd}, the viratual state~\cite{Padmanath:2022cvl,Lyu:2023xro,Dai:2021wxi}, or the Efimov state~\cite{Liu:2019yye}, etc.

From the experimental observations, most exotic states contain heavy flavor quarks, while research on light flavor quark system is comparatively sparse in both experimental and theoretical aspects. Given this, if quark mass differences under an appropriate approximation are neglected,  the discovery of $T_{cc}^{+}(3875)$ state raises the question of whether similar double-strange tetraquark states could also exist. This intriguing possibility deserves further exploration.  These double-strange tetraquark states are of comparable importance to the double-charm tetraquark $T_{cc}^{+}(3875)$, as they offer essential insights into the formation mechanisms of tetraquark states in the light quark sector.

Although theoretcal studies on double-strange tetraquark states are more restricted at present, there have been some preliminary advances. For example, in Ref.~\cite{Wang:2024kke}, the spectral properties of the $\bar{K}^{\ast}K^{\ast}$ system have been investigated by means of the one-boson-exchange model, and it is suggested that $\bar{K}\bar{K}^{\ast}$ and $\bar{K}^{\ast}\bar{K}^{\ast}$ with $I(J^{P})=0(1^{+})$ may be candidates of the double-strange tetraquark state. Similar double-strange tetraquark states have also been explored in Ref.~\cite{Ji:2024znj}. Additionaly, Refs.~\cite{Beane:2007uh,Kanada-Enyo:2008wsu} utilized lattice QCD to calculate the scattering phase shifts and effective potential for the $I=1$ system, finding the effective interaction to be repulsive. The mutual exclusion of $\bar{K}\bar{K}$ had also been verified in the investigation of systems $\bar{K}\bar{K}N$~\cite{Shevchenko:2015oea,Kezerashvili:2016eic} and $\bar{K}\bar{K}NN$~\cite{Marri:2016qsf,Barnea:2012qa}.

This paper aims to systematically examine the spectral properties of double-strange tetraquark systems. Utilizing QDCSM, we perform dynamical calculations to evaluate the effective interactions within these systems and solve the Schrödinger equation to identify prospective double-strange tetraquark states. Moreover, we apply the real-scaling method to predict resonance states within these systems. It is expected that the results of this study will provide valuable guidance for experimental efforts directed toward the discovery of double-strange tetraquark states, thereby advancing the broader investigation of tetraquark states in the light quark sector.

The paper is arranged as follows.  In the next section, the theoretical framework of QDCSM model and  the resonating group method are briefly introduced. The calculated resultes are listed in Section~\ref{results}, where some discussions are made as well. Finally, the summary is given in Section~\ref{Sum}.

\section{Quark delocalization color screening model and  the resonanting group method }{\label{model}}

\subsection{Quark delocalization color screening model}
The QDCSM is an extension of the native quark cluster model~\cite{DeRujula:1975qlm,Isgur:1978xj,Isgur:1978wd,Isgur:1979be} and was also developed with aim of addressing multiquark systems. For the tetraquark system, the Hamiltonian reads,
\begin{equation}
H = \sum_{i=1}^{4} \left(m_i+\frac{\boldsymbol{p}_i^2}{2m_i}\right)-T_{CM}+\sum_{j>i=1}^4V(r_{ij}),\\
\end{equation}
where $T_{CM}$ is the center-of-mass kinetic energy, who is usually subtracted without losing generality since we mainly focus on the internal relative motions of the multiquark system. The interplay is of two body potential, which includes color-confining term $V_{\mathrm{CON}}$, one-gluon exchange potential $V_{\mathrm{OGE}}$, and the potential results from Goldstone-boson exchange, $V_{\chi}$, i.e.,
\begin{equation}
V(r_{ij}) = V_{\mathrm{CON}}(r_{ij})+V_{\mathrm{OGE}}(r_{ij})+V_{\chi}(r_{ij}).
\end{equation}

In the present work, we focus on the $S-$wave low-lying positive  $ss\bar{q}\bar{q}$ tetraquark system with positive parity where q represents u or d. In this case, the spin-orbit and tensor interactions vanish and the potential $V_{\mathrm{OGE}}(r_{ij})$ become,
\begin{eqnarray}
\nonumber
V_{\mathrm{OGE}}(r_{ij}) &=& \frac{1}{4}\alpha_s^{q_{i}q_{j}} \boldsymbol{\lambda}^{c}_i \cdot
\boldsymbol{\lambda}^{c}_j \\
&&\left[\frac{1}{r_{ij}}-\frac{\pi}{2}\delta(\boldsymbol{r}_{ij})(\frac{1}{m^2_i}+\frac{1}{m^2_j}
+\frac{4\boldsymbol{\sigma}_i\cdot\boldsymbol{\sigma}_j}{3m_im_j})\right],
\end{eqnarray}
where $m_{i}$ is the quark mass,  $\boldsymbol{\sigma}_i$ and $\boldsymbol{\lambda^{c}}_i$ are the Pauli matrices and SU(3) color matrices, respectively. The $\alpha_s^{q_{i}q_{j}}$ is the quark-gluon coupling constant, which offers a consistent description of mesons from light to heavy-quark sector.The values of $\alpha_{ij}$ are associated with the quark flavors and in the present work they are fixed by reproducing the mass difference of the low-lying mesons with $S=0$ and $S=1$.

The confining potential $V_{\mathrm{CON}}(r_{ij})$ is
\begin{equation}
 V_{\mathrm{CON}}(r_{ij}) =  -a_{c}\boldsymbol{\lambda^{c}_{i}\cdot\lambda^{c}_{j}}\left[f(r_{ij})+V_{0_{q_{i}q_{j}}}\right],
\end{equation}
where the $V_{0_{q_{i}q_{j}}}$ is determined by the mass shift of the theoretical esmations and experimental measurement of each kind of meson, which is also quark flavor related parameter. In the QDCSM, the function $f(r_{ij})$ is,
\begin{equation}
 f(r_{ij}) =  \left\{ \begin{array}{ll}r_{ij}^2  &\qquad \mbox{if }i,j\mbox{ occur in the same cluster} , \\
\frac{1 - e^{-\mu_{ij} r_{ij}^2} }{\mu_{ij}} & \qquad \mbox{if }i,j\mbox{ occur in different cluster} , \\
\end{array} \right.
\end{equation}
where the color screening parameter $\mu_{ij}$ relevant to the light quarks can be determined by fitting the deuteron properties, $NN$ and $NY$ scattering phase shifts~\cite{Chen:2011zzb,Ping:1993me,Wang:1998nk}.

The Goldstone-boson exchange interactions between light quarks appear because the dynamical breaking of chiral symmetry. For the $\ssqq$ system,  the concrete form of the Goldstone-boson exchange potential becomes,
\begin{eqnarray}
V^{\chi}_{ij}  &=& V_{\pi}(\boldsymbol{r}_{ij})\sum_{a=1}^3\lambda
_{i}^{a}\cdot \lambda _{j}^{a}+V_{K}(\boldsymbol{r}_{ij})\sum_{a=4}^7\lambda
_{i}^{a}\cdot \lambda _{j}^{a}+ \nonumber\\
&&V_{\eta}(\boldsymbol{r}_{ij})\left[\left(\lambda _{i}^{8}\cdot
\lambda _{j}^{8}\right)\cos\theta_P-(\lambda _{i}^{0}\cdot
\lambda_{j}^{0}) \sin\theta_P\right], \label{sala-Vchi1}
\end{eqnarray}
with
\begin{eqnarray}
\nonumber
V_{\chi}(\boldsymbol{r}_{ij}) &=&  {\frac{g_{ch}^{2}}{{4\pi}}}{\frac{m_{\chi}^{2}}{{\
12m_{i}m_{j}}}}{\frac{\Lambda _{\chi}^{2}}{{\Lambda _{\chi}^{2}-m_{\chi}^{2}}}}
m_{\chi}                                 \\
&&\left\{(\boldsymbol{\sigma}_{i}\cdot\boldsymbol{\sigma}_{j})
\left[ Y(m_{\chi}\,r_{ij})-{\frac{\Lambda_{\chi}^{3}}{m_{\chi}^{3}}}
Y(\Lambda _{\chi}\,r_{ij})\right] \right\},\nonumber \\
&& ~~~~~~\chi=\{\pi, K, \eta\},
\end{eqnarray}
where $Y(x)=e^{-x}/x$ is the standard Yukawa function. The $\boldsymbol{\lambda^{a}}$ is the SU(3) flavor Gell-Mann matrix. The masses of the $\pi$, $K$ and  $\eta$ meson are taken from the experimental value~\cite{ParticleDataGroup:2022pth}, respectively. The chiral coupling constant, $g_{ch}$, is determined from the $\pi NN$ coupling constant through,
\begin{equation}
\frac{g_{ch}^{2}}{4\pi}=\left(\frac{3}{5}\right)^{2} \frac{g_{\pi NN}^{2}}{4\pi} {\frac{m_{u,d}^{2}}{m_{N}^{2}}},
\end{equation}
where the SU(3) flavor symmetry only broken by the different masses of the light quarks. All the other model parameters are  the same as the ones in Ref.~\cite{Xue:2020vtq}, which were determined by reproducing the mass spectrum of the ground  mesons. The authors of Ref.~\cite{Xue:2020vtq} used the same model to investigate the tetraquarks with open charm flavor and found that the $X_{0}(2900)$ could be interpreted as the molecular state $\bar{D}K^{\ast}$ with $I(J^{P})=0(0^{+})$.

In the QDCSM, the single-particle orbital wave functions in the ordinary quark cluster model are the left and right centered single Gaussian functions, which are,
\begin{eqnarray}\label{single}
\phi_\alpha(\boldsymbol {S_{i}})=\left(\frac{1}{\pi
b^2}\right)^{\frac{3}{4}}e^ {-\frac{(\boldsymbol {r_{\alpha}}-\frac{1}{2}\boldsymbol
{S_i})^2}{2b^2}},
 \nonumber\\
\phi_\beta(-\boldsymbol {S_{i}})=\left(\frac{1}{\pi
b^2}\right)^{\frac{3}{4}}e^ {-\frac{(\boldsymbol {r_{\beta}}+\frac{1}{2}\boldsymbol
{S_i})^2}{2b^2}} .
 \
\end{eqnarray}
The quark delocalization is realized by writing the single-particle orbital wave function as a
linear combination of the left and right Gaussians, which are,
\begin{eqnarray}
{\psi}_{\alpha}(\boldsymbol {S_{i}},\epsilon) &=&
\left({\phi}_{\alpha}(\boldsymbol{S_{i}})
+\epsilon{\phi}_{\alpha}(-\boldsymbol{S_{i}})\right)/N(\epsilon),
\nonumber \\
{\psi}_{\beta}(-\boldsymbol {S_{i}},\epsilon) &=&
\left({\phi}_{\beta}(-\boldsymbol{S_{i}})
+\epsilon{\phi}_{\beta}(\boldsymbol{S_{i}})\right)/N(\epsilon),
\nonumber \\
N(\epsilon)&=&\sqrt{1+\epsilon^2+2\epsilon e^{{-S}_i^2/4b^2}}.
\end{eqnarray}
where $\epsilon(\boldsymbol{S}_i)$ is the delocalization parameter determined by the dynamics of the quark system rather than free parameters. In this way, the system can choose its most favorable configuration through its dynamics in a larger Hilbert space.

\begin{figure}[htb]
    \includegraphics[scale=0.65]{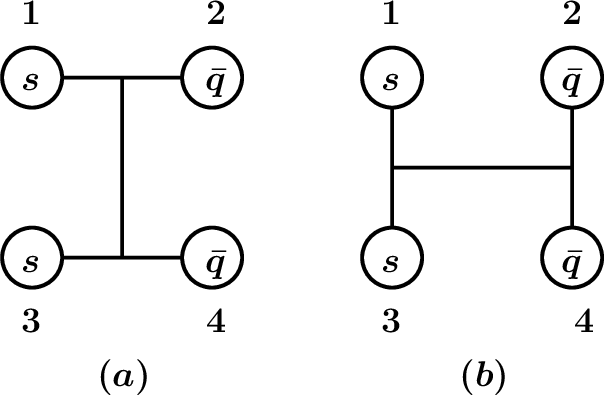}
    \caption{The meson-meson configuration [diagram (a)] and diquark-antidiquark configuration [diagram (b)]  in the $\ssqq$ tetraquark system."$\bar{q}$" represents $\bar{u}$, and $\bar{d}$.}
    \label{fig1}
\end{figure}

\subsection{The resonating group method}
In the present work, the RGM is employed to carry out the dynamical calculation. The core of this approach is how to deal with the two-body problem~\cite{Kamimura:1981oxj,Kamimura:1977okl}. When dealing with the two-cluster system in this method, one can only consider the relative motion between the clusters, while the two clusters are frozen inside. So the wave function of the $\ssqq$ system can be constructed as,
\begin{eqnarray}\label{wave1}
  \psi_{4q} &=&  \mathcal{A}\left[\left[\psi_{A}(\boldsymbol{\rho}_A)\psi_{B}(\boldsymbol{\rho}_B)\right]^{[\sigma]IS}\otimes\chi_{L}(\textbf{R})\right]^{J},
\end{eqnarray}
where the symbol $\mathcal{A}$ is the antisymmetry operator, which is defined as
\begin{eqnarray}\label{wave2}
    \mathcal{A}&=& 1-P_{13}-P_{24}+P_{13}P_{24}.
\end{eqnarray}
where the $P_{ij}$ indicates the exchange of the particle positions with numbers $i$ and $j$ from the Fig.~\ref{fig1}. $[\sigma]=[222]$ gives the total color symmetry. The symbols $I$, $S$, $L$, and $J$ represent flavor, spin, orbit angular momentum, and total angular momentum of $\ssqq$ system, respectively. $\psi_{A}$ and $\psi_{B}$ are the wave functions of the two-quark cluster, which are,
\begin{eqnarray}
 \psi_{A} &=& \left(\frac{1}{2\pi b^{2}}\right)^{3/4}  e^{-\boldsymbol{\rho_{A}}^{2}/(4b^2)} \eta_{I_{A}}S_{A}\chi_{A}^{c} ,\nonumber\\
  \psi_{B} &=& \left(\frac{1}{2\pi b^{2}}\right)^{3/4} e^{-\boldsymbol{\rho_{B}}^{2}/(4b^2)} \eta_{I_{B}}S_{B}\chi_{B}^{c} ,
\end{eqnarray}
where $\eta_I$, $S$, and $\chi$ represent the flavor, spin and internal color terms of the cluster wave functions, respectively. According to Fig.~\ref{fig1}, we adopt different Jacobi coordinates for different diagrams. As for the meson-meson configuration in Fig.~\ref{fig1}-(a), the Jacobi coordinates are defined as,
\begin{eqnarray}\label{wave6}
\boldsymbol{\rho_{A}}&=&{\br_{q_1}-\br_{\bar{q}_2}},  \ \ \ \ \boldsymbol{\rho_{B}}={\br_{q_3}-\br_{\bar{q}_4}},\nonumber \\
\boldsymbol{R_{A}}&=& \frac{m_{1}{\br_{q_1}}+m_{2} {\br_{\bar{q}_2}}}{m_{1}+m_{2}},\nonumber \\
 \boldsymbol{R_{B}}&=& \frac{m_{3}{\br_{q_3}}+m_{4}{\br_{\bar{q}_4}}}{m_{3}+m_{4}},\nonumber \\
\boldsymbol{R}&=&\boldsymbol{R_{A}-R_{B}}, \nonumber \\
\boldsymbol{R_{c}}&=&\frac{m_{1}{\br_{q_1}}+m_{2}{\br_{\bar{q}_2}}+m_{3}{\br_{q_3}}+m_{4}{\br_{\bar{q}_4}}}{m_{1}+m_{2}+m_{3}+m_{4}}.
 % \rho_{\bf A} &=& \bf r_{1}- \bf r_{2},  \rho_{\bf B} =  \bf r_{3}- \bf r_{4}, \\
 % \bf R_{A} &=& \frac{1}{2}(\bf r_{1}}+\bf r_{2}),  \boldsymbol{R_{B}} = \frac{1}{2}(r_{3}+r_{4}),\\
%  \textbf{R} &=& \textbf{R_{A}}-\textbf{R_{B}}, \textbf{R_{c}}=\frac{1}{2}(\textbf{R_{A}}+\textbf{R_{B}}).
\end{eqnarray}
where the subscript $q/\bar{q}$ indicates the quark or antiquark particle, while the number indicates the quark position in Fig.~\ref{fig1}-(a). As for the diquark-antidiquark configuration as shown in Fig.~\ref{fig1}-(b), the relevant Jacobi coordinates can be obtained by interchanging $\br_{q_3}$ with $\br_{\bar{q}_2}$ in Eq.~(\ref{wave6}).

Form the variational principle, after variation with respect to the relative motion wave function $\chi\boldsymbol(R)=\sum_{L}\chi_{L}\boldsymbol(R)$, one obtains the RGM equation, which is,
\begin{eqnarray}\label{wave3}
  \int H\left(\boldsymbol{R, R^{\prime}}\right)\chi\left(\boldsymbol{R^{\prime}}\right)d\boldsymbol\left(R^{\prime}\right)=E\nonumber \\  \int N\left(\boldsymbol{R, R^{\prime}}\right) \chi\left(\boldsymbol{R^{\prime}}\right)d\boldsymbol\left(R^{\prime}\right),
\end{eqnarray}
where $H(\boldsymbol{R, R^{\prime}})$ and $N(\boldsymbol{R, R^{\prime}})$    are Hamiltonian and norm kernels, respectively. The eigenenergy $E$ and the wave functions can be obtained by solving the RGM equation. In the present estimation, the function $\chi (\boldsymbol{R})$ can be expanded by gaussian bases,  which is
\begin{eqnarray}\label{wave5}
\chi\boldsymbol{(R)}&=&\frac{1}{\sqrt{4\pi}}\sum_{L}\left(\frac{1}{\pi b^2}\right)^{3/4}\sum_{i}^{n}C_{i,L} \nonumber\\
&&\times\int e^{-\frac{1}{2}\boldsymbol(R-S_{i})^{2}/b^2} Y^L\left(\hat{\boldsymbol{S}_{i}}\right)d\hat{\boldsymbol{S}_{i}},
\end{eqnarray}
where $C_{i,L}$ is the expansion coefficient, and $n$ is the number of gaussian bases, which is determined by the stability of the results. $\boldsymbol{S}_{i}$ is the separation of two reference centers. $\boldsymbol{R}$ is the dynamic coordinate defined in Eq.~(\ref{wave6}). After including the motion of the center of mass, i.e., %In each cluster, the reference center is fixed, and the quarks move around the reference center, whereas the dynamic coordinate $\boldsymbol{R}$ is a quantity varies with the motion of each quark.
\begin{equation}
  \phi_{C}(\boldsymbol{R_{c}})=\left(\frac{4}{\pi b^2}\right)^{3/4}\mathrm{e}^{\frac{-2 \boldsymbol{R_{c}}^{2}}{b^{2}}},
\end{equation}
one can rewrite Eq.~(\ref{wave2}) as,
\begin{eqnarray}\label{wave4}
%\begin{split}
\psi_{4q}&=&\mathcal{A} \sum_{i,L}C_{i, L}\int \frac{d \hat{\boldsymbol{S_{i}}}}{\sqrt{4 \pi}} \prod_{\alpha=1}^{2}\phi_{\alpha}\left(\boldsymbol{S_{i}}\right)\prod_{\alpha=3}^{4}\phi_{\beta}\left(\boldsymbol{-S_{i}}\right) \nonumber \\
&&\times \left[\left[\eta_{I_{A}S_{A}}\eta_{I_{B}S_{B}}\right]^{IS}Y^{L}(\hat{\boldsymbol{S_{i}}})\right]^{J}\left[\chi_{A}^{c}\chi_{B}^{c}\right]^{[\sigma]} ,
%\end{split}
\end{eqnarray}
where $\phi_{\alpha}(\boldsymbol{S_{i}})$ and $\phi_{\beta}(\boldsymbol{-S_{i}})$ are the single-particle orbital wave functions with different reference centers, whose specific expressions have been presented in Eq.~(\ref{single}).

With the reformulated ansatz as shown in Eq.~(\ref{wave4}), the RGM equation becomes an algebraic eigenvalue equation, which is,
\begin{eqnarray}
  \sum_{j,L}C_{J,L}H_{i,j}^{L,L^{\prime}} &=& E\sum_{j}C_{j,L^{\prime}}N_{i,j}^{L^{\prime}},
\end{eqnarray}
where $N_{i,j}^{L^{\prime}}$ and $H_{i,j}^{L,L^{\prime}}$ are the  overlap of the wave functions and the matrix elements of the Hamiltonian, respectively. By solving the generalized eigenvalue problem, we can obtain the energies of the tetraquark  systems $E$ and the corresponding expansion coefficients $C_{j,L}$. Finally, the relative motion wave function between two clusters can be obtained by substituting the $C_{j,L}$ into Eq.~(\ref{wave5}). As for the flavor, spin and color wave functions of the tetraquark system, they are constructed in a two step way. One can first construct the wave functions for the two clusters, and then coupling the wave functions of two clusters to form the wave function for tetraquark system. The details of the flavor, spin and color wave function for tetraquark system are collected in the Appendix \ref{Sec:App}.

 In addition, it is noteworthy that the quark composition of the $\ssqq$ system exhibits uniqueness, as its structure includes two identical quarks and two identical antiquarks. Therefore, the wave function of the $\ssqq$
 tetraquark states, constructed from four degrees of freedom (orbital, color, spin, flavor), should satisfy antisymmetry requirements. Since  the present study consider only the $S-$wave case, under this constraint, the wave functions constructed for quantum numbers $I(J^{P})=0(0^{+})$,  $0(2^{+})$, and $1(1^{+})$ with S=1 $(S_{1}\otimes S_{2}=1\otimes1)$ fail to meet the antisymmetry requirements and must therefore be excluded in the $\ssqq$ tetraquark systems. Tables~\ref{channels} and ~\ref{channels} summarize the  construction of color, spin, and flavor bases for all possible quantum numbers.

\begin{table}[t]
    \begin{center}
        \caption{\label{channels} The  constructed relevant channels  for all possible quantum numbers. Squark brackets [i,j,k] represents the flavor, spin, and color wave functions corresponding to the relevant channels, where "i", "j", and "k" respectively denote the selected flavor, spin, and color wave functions.  }
        \renewcommand\arraystretch{1.2}
        %\resizebox{0.48\textwidth}{!} {
        \begin{tabular}{m{1.5cm}<\centering m{1.cm}<\centering m{2.5cm}<\centering m{1.5cm}<\centering }%{18cm}{@{\extracolsep{\fill}}*{11}{p{1.3cm}<{\centering}}}
        \toprule[1pt]
         $I(J^{P})$  &  Index &  $F^{i}; S_{s}^{j};\chi_{k}^{c}$ [i,j,k] & Channels \\
           \midrule[1pt]
       \multirow{5}{*}{$0(1^{+})$} & 1 & [1,3,1] & $\bar{K}^{0} K^{\ast-}$    \\
                                                                        & 2 & [1,4,1] & $\bar{K}^{\ast0} K^{-}$   \\
                                                                        & 3 & [1,5,1] & $\bar{K}^{\ast0} K^{\ast-}$    \\
                                                                        & 4 & [3,3,3] & $(ss)(\bar{n}\bar{n})$    \\
                                                                        & 5 & [4,4,3] & $(ss)(\bar{n}\bar{n})$     \\
        \midrule[1pt]
       \multirow{4}{*}{$1(0^{+})$} & 1 & [1,1,2] & $\bar{K}^{0} \bar{K}^{0}$   \\
                                                                  & 2 & [1,2,2] & $\bar{K}^{\ast0} \bar{K}^{\ast0}$   \\
                                                                  & 3 & [3,1,4] & $(ss)(\bar{n}\bar{n})$   \\
                                                                  & 4 & [4,2,4] & $(ss)(\bar{n}\bar{n})$   \\                                                                 
         \midrule[1pt]                                                         
        \multirow{3}{*}{$1(1^{+})$} & 1 & [1,3,2] & $\bar{K}^{0} \bar{K}^{\ast0}$   \\
                                                                   & 2 & [1,4,2] & $\bar{K}^{\ast0} \bar{K}^{0}$   \\   
                                                                   & 3 & [4,5,4] & $(ss)(\bar{n}\bar{n})$\\   
                                                                                             
          \midrule[1pt]                                                         
         \multirow{2}{*}{$1(2^{+})$} & 1 & [1,6,2] &$\bar{K}^{\ast0} \bar{K}^{\ast0}$   \\
                                                                    & 2 & [4,6,4] & $(ss)(\bar{n}\bar{n})$ \\
                                                                                                                     
        \bottomrule[1pt]

        \end{tabular}
 %   }
    \end{center}
\end{table}

\section{Numerical RESULTS AND DISCUSSIONS}{\label{results}}
In the present work,  the investigation of the doubly strange tetraquark states under low-energy $S-$wave scenario  is explored. As shown in Fig.~\ref{fig1},  the two structural possibilities considered are meson-meson and diquark-antidiquark configurations.  The allowed quantum numbers are $I(J^{P})=0(1^{+})$, $1(0^{+})$, $1(1^{+})$ and $1(2^{+})$ under the constraints of the Pauli principle for the $\ssnn$ tetraquark states .  In the present estimations, to assess whether bound states or resonance states exist in the doubly strange tetraquark system, the resonance group method is utilized to evaluate the existence of bound states for different quantum numbers, with the real scaling method applied to search for possible resonance states.

%while the allowed quantum numbers in the $\ssss$ tertraquark states are $I(J^{P})=0(0^{+})$, $0(1^{+})$, and $0(2^{+})$ as the isospin of s-quark is 0

\begin{table}[!htb]
\begin{center}
\caption{\label{ssnn-bound} The lowest-lying eigenenergies (in unit of MeV) of the $\ssnn$ tetraquark states with all possible quantum numbers. }
\renewcommand\arraystretch{1.3}
%\resizebox{0.48\textwidth}{!} {
\begin{tabular}{p{1.2cm}<\centering p{1.cm}<\centering p{1.5cm}<\centering p{1.cm}<\centering p{1.cm}<\centering p{1.cm}<\centering p{1.cm}<\centering p{1.cm}<\centering p{1.cm}<\centering p{0.5cm}<\centering p{0.5cm}<\centering p{1.cm}<\centering p{1.cm}<\centering p{1.2cm}<\centering p{1.5cm}<\centering }%{18cm}{@{\extracolsep{\fill}}*{11}{p{1.3cm}<{\centering}}}
\toprule[1pt]
 $I(J^{P})$  &  Index &   Channels  & $E_{th}$ & $E_{sc}$ & $E_{cc}$ &$E_{mix}$\\
\midrule[1pt]
 \multirow{5}{*}{$0(1^{+})$}  &1  &$\bar{K}^{0} K^{\ast-}$               &1386.2  & 1384.7 &1324.8 &1309.7 \\
                                                 &2  &$\bar{K}^{\ast0} K^{-}$          &1386.2  &1384.7  \\
                                                 &3  & $\bar{K}^{\ast0} K^{\ast-}$   &1784.2  &1779.8 \\
                                                 &4  & $(ss)(\bar{n}\bar{n})$             &            &1818.0 & 1415.8\\
                                                 &5  & $(ss)(\bar{n}\bar{n})$             &            &1513.6\\                                               
\midrule[1pt]
\multirow{4}{*}{$1(0^{+})$}  &1  & $\bar{K}^{0} \bar{K}^{0}$    &988.2    & 992.4 &992.4. &991.7 \\
&2  &$\bar{K}^{\ast0} \bar{K}^{\ast0}$                                          &1784.2  &1787.3  \\
&3  & $(ss)(\bar{n}\bar{n})$                                                             &            &1982.1 &1391.3\\
&4  & $(ss)(\bar{n}\bar{n})$                                                             &            &1711.9\\                                               
\midrule[1pt]
 \multirow{3}{*}{$1(1^{+})$}  &1  &$\bar{K}^{0} \bar{K}^{\ast0}$     &1386.2 &1390.5 &1390.5 &1390.2 \\
&2  &$ \bar{K}^{\ast0} \bar{K}^{0}$                                                     &1386.2 &1390.5  \\
&3  & $(ss)(\bar{n}\bar{n})$                                                                   &           &1812.1\\                                               
\midrule[1pt]
 \multirow{2}{*}{$1(2^{+})$}  &1  &$\bar{K}^{\ast0} \bar{K}^{\ast0}$             &1784.2 & 1786.9 &1786.9 &1786.4 \\
&2  & $(ss)(\bar{n}\bar{n})$                                                                            &           &1964.2\\                                               
\midrule[1pt]
\end{tabular}
%}
\end{center}
\end{table}

\subsection{The Bound  States }
For the tetraquark system containing $\ssnn$, a systematic theoretical estimation of bound states is performed. The estimation results for the $\ssnn$ tetraquark states under different quantum numbers are summarized in Table~\ref{ssnn-bound}. As shown in the table, the results include not only single-channel estimates for each configuration but also channel coupling estimates for the two configurations. The first column, labeled "Index", represents the symbols for each channel, while the second and third columns list all constructed relevant channels and their theoretical thresholds, respectively. Additionally, the fourth column, denoted as $E_{sc}$, represents the lowest eigenenergy obtained in the single-channel estimation, whereas $E_{cc}$ and $E_{mix}$ are the lowest estimated eigenenergies obtained by considering the coupled channel effects of each kind of configuration and both configurations, respectively.

 Additionally, it is worth noting that the binding energy of the $\ssnn$ tetraquark state, $E_{b}$, is defined as $E_{b}=E_{i}-E_{4}(\infty)$, to evaluate the stability of the $\ssnn$ tetraquark state under strong interactions. Here, $E_{4}(\infty)$ represents the lowest threshold of the two-meson configuration estimated within the QDCSM, and $i$ denotes different channel coupling scenarios. This definition significantly reduces the impact of the model parameters on the binding energy. If $E_{b}\geq 0$, the tetraquark system can decay into two mesons via strong interactions. Conversely, if $E_{b}\leq0$, the strong decay into two mesons is kinematically forbidden, and the decay of the tetraquark state can only occur through weak or electromagnetic interactions.

 For the system with $I(J^{P})=0(1^{+})$,  the estimated eigenenergies of five channels including three channels in meson-meson configurations and two channels in diquark-antidiquark configurations are presented in Table~\ref{ssnn-bound}. For the meson-meson configuration, the single-channel estimations show that the lowest eigenenergies for the $\bar{K}^{0}K^{\ast-}$, $\bar{K}^{\ast0}K^{-}$, and $\bar{K}^{\ast0}K^{\ast-}$  channels are all below their respective theoretical thresholds, indicating they form bound states. Their binding energies are approximately $-1.5$ MeV, $-1.5$ MeV,  and $-4.4$ MeV, respectively. When considering the channel coupling effects in the meson-meson configurations,  the bound state with binding energy approximately $-61.4$ MeV can be obtained, which indicates that the coupled channel effect in the meson-meson configuration is rather strong.  For the diquark-antidiquark configuration,  the lowest eigenenergies obtained from single-channel estimations are above the threshold of the lowest meson-meson channel ($\bar{K}^{0}K^{\ast-}$). Similarly, the lowest eigenenergy obtained from the coupled-channel estimation within the diquark-antidiquark sector also lies above this $\bar{K}^{0}K^{\ast-}$ threshold.  However, the channel coupling estimation within the diquark-antidiquark configuration yields the lowest eigenenergy of approximately  $-100$ MeV less than the lowest single-channel theoretical threshold, highlighting the significant role of the color structure's channel coupling effects. 
 Moreover, the evaluation of the coupling between the meson-meson configuration and diquark-antidiquark configuration demonstrates that their interaction facilitates the formation of a bound state in the $\ssnn$ tetraquark system, with a binding energy of roughly $-76.5$ MeV.  This estimated result is similar to the channel coupling results reported in Refs~\cite{Wang:2024kke,Ji:2024znj}, all indicating the existence of a $\bar{K}^{0}K^{\ast-}$ bound state.

For the system with $I(J^{P})=1(0^{+})$,  there are four channels, which are $\bar{K}^{0}\bar{K}^{0}$, $\bar{K}^{*0}\bar{K}^{*0}$, and two color structures with $\textbf{6}_{c}\otimes\bar{\textbf{6}}_{c}$ and $\bar{\textbf{3}}_{c}\otimes\textbf{3}_{c}$ in the diquark-antidiquark configuration. From Table~\ref{ssnn-bound}, it can be seen that the single-channel estimated eigenenergies are all above the corresponding threshold in the meson-meson configuration, indicating that the single channel estimations do not support the existence of the bound states. Additionally, the coupling of channels $\bar{K}^{0}\bar{K}^{0}$ and $\bar{K}^{*0}\bar{K}^{*0}$ in the meson-meson configuration resulted in a minimum eigenenergy estimate of 992.4 MeV, which is still above the theoretical threshold for channel $\bar{K}^{0}\bar{K}^{0}$. It implies that the significant mass difference between the two channels leads to diminished coupling effects. A similar phenomenon occurs in the diquark-antidiquark configuration, as both single channel and channel coupling estimations yield the lowest eigenenergies surpassing the theoretical threshold of $\bar{K}^{0}\bar{K}^{0}$.  Comparing the channel coupling estimates of the meson-meson and diquark-antidiquark configurations reveals that the coupling effects of the color structure are more pronounced than that of the meson-meson configuration. Furthermore, the coupling estimates of both configurations, as illustrated in Table~\ref{ssnn-bound},  provide the lowest eigenenergy of 991.7 MeV, still exceeding the lowest theoretical threshold of 988.2 MeV, which suggests no bound state exists in the $\ssnn$ tetraquark system with $I(J^{P})=1(0^{+})$.

For the system with $I(J^{P})=1(1^{+})$,  the lowest eigenenergies derived from single-channel estimations within both the meson-meson and diquark-antidiquark configurations exceed the $\bar{K}^{0}\bar{K}^{\ast 0}$ threshold. Additionally, when considering the full channel coupling between the meson-meson and diquark-antidiquark configurations, the lowest resulting eigenenergy is approximately 1390.2 MeV, which is significantly higher than the $\bar{K}^{0}\bar{K}^{\ast 0}$ channel threshold of 1386.2 MeV. These findings suggest that a bound state does not exist in the $\ssnn$ tetraquark system with $I(J^{P})=1(1^{+})$. Analogous conclusions can be drawn for the $\ssnn$ tetraquark system with $I(J^{P})=1(2^{+})$. In this case, all estimations, including channel coupling within meson-meson configurations, within diquark-antidiquark configurations, and the full coupling between both types, yield the lowest eigenenergies that consistently exceed the minimum threshold of $\bar{K}^{\ast 0} \bar{K}^{\ast 0}$. A comparison with estimation results from the OBE model for systems with quantum number $I=1$ reveals nearly identical conclusions. Both this study and the OBE model find no bound states in these channels~\cite{Wang:2024kke}.

\begin{figure}[htb]
    \includegraphics[scale=0.2]{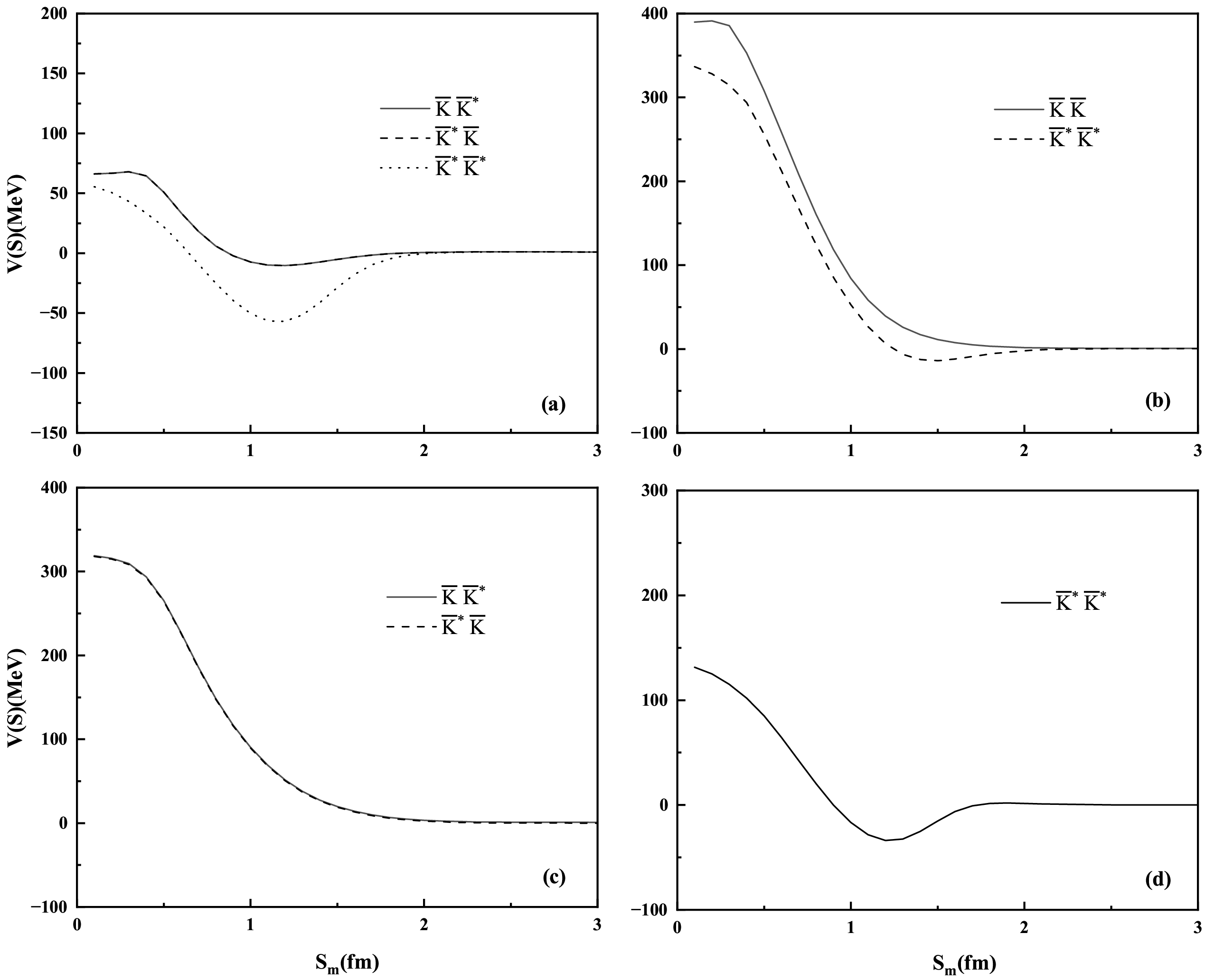}
    \caption{The effective potential for the $\ssnn$ system with meson-meson configuration. (a) represents the system with $I(J^{P})=0(1^{+})$; (b) denotes the system with $I(J^{P})=1(0^{+})$; (c) stands for the system with $I(J^{P})=1(1^{+})$; (d) is the system with $I(J^{P})=1(2^{+})$ .  }
    \label{meson-effective}
\end{figure}

\begin{figure}[htb]
    \includegraphics[scale=0.2]{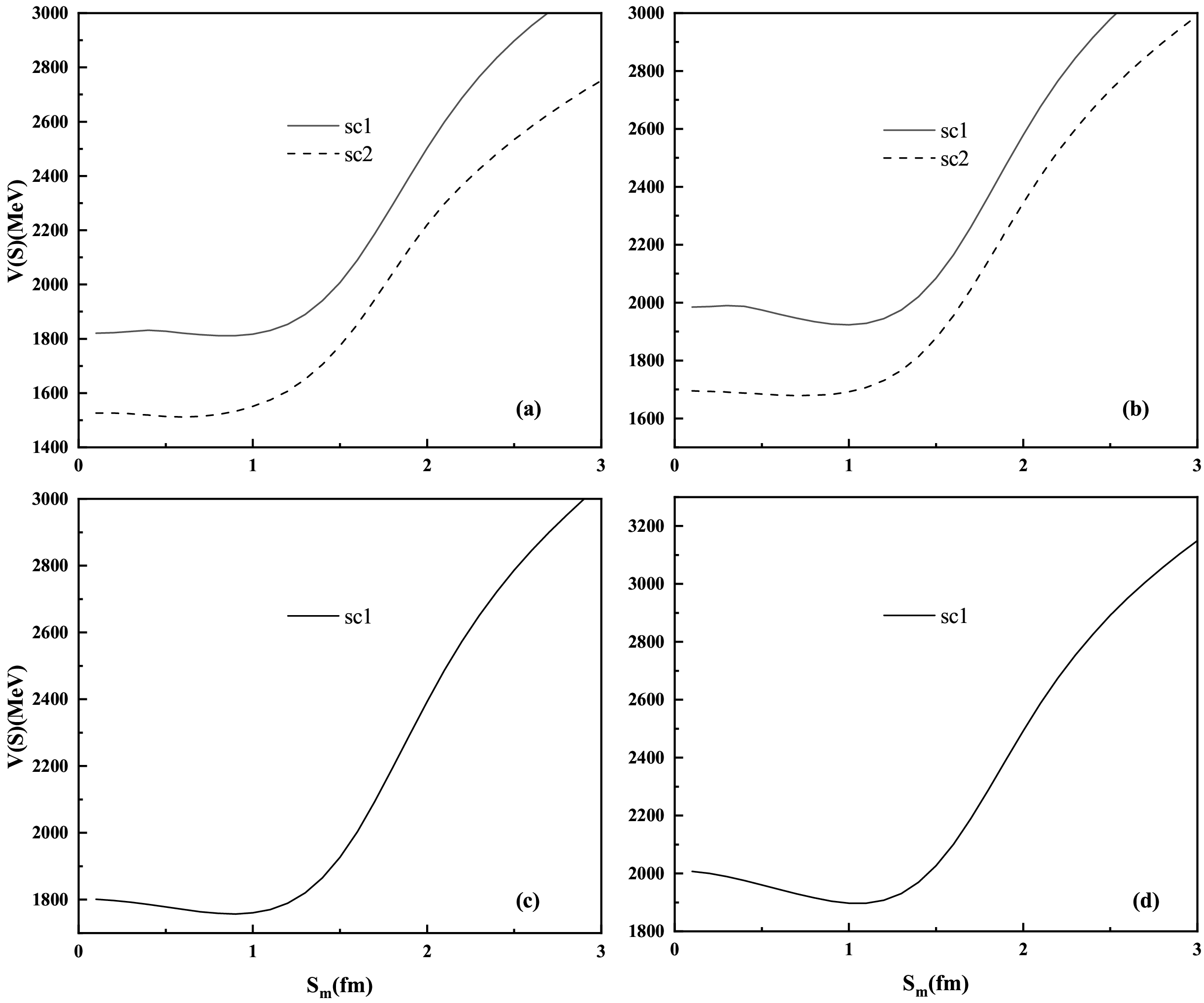}
    \caption{The effective potential for the $\ssnn$ system with diquark-antiquark configuration. (a) represents the system with $I(J^{P})=0(1^{+})$; (b) denotes the system with $I(J^{P})=1(0^{+})$; (c) stands for the system with $I(J^{P})=1(1^{+})$; (d) is the system with $I(J^{P})=1(2^{+})$ . sc$i$ indicates the effective potential of $i-$th diquark-antidiquark channel .}
    \label{diquark-effective}
\end{figure}

The conclusions drawn from the bound state estimation prompt a more rigorous investigation into the interactions between the two clusters, with the aim of determining whether a mechanism for bound state formation exists. For the $\ssnn$ tetraquark system,  Figures~\ref{meson-effective} and~\ref{diquark-effective} present the equivalent potentials of the meson-meson and diquark-antidiquark configurations under various quantum numbers.  In the meson-meson configuration, when the quantum number is $0(1^{+})$, as shown in  Fig.~\ref{meson-effective}(a), the equivalent potentials for channels $\bar{K}^{0} K^{\ast-}$, $\bar{K}^{\ast0} K^{-}$ and $\bar{K}^{\ast0} K^{\ast-}$ exhibit attractive interactions. This clearly explains why these channels can form bound states in the single-channel estimation. Notably, bound states form even though the attractive interactions for $\bar{K}^{0} K^{\ast-}$ and $\bar{K}^{\ast0} K^{-}$  are weaker than those for $\bar{K}^{\ast0} K^{\ast-}$.   From Figs.~\ref{meson-effective}(b) and (c), it can be observed that  the interactions for the $\bar{K}^{0}\bar{K}^{0}$ with $1(0^{+})$, $\bar{K}^{0}\bar{K}^{\ast0}$ and  $\bar{K}^{\ast0}\bar{K}^{0}$ with $1(1^{+})$ are repulsive. Furthermore, the repulsive interaction between $\bar{K}^{0}\bar{K}^{0}$ with $1(0^{+})$ has been consistently verified in lattice QCD calculations and the OBE model~\cite{Wang:2024kke,Beane:2007uh,Kanada-Enyo:2008wsu}. Therefore, bound states cannot form in these channels within single-channel estimations. In Fig.~\ref{meson-effective}(b) and Fig.~\ref{meson-effective}(d), although the interactions of the remaining channels $\bar{K}^{\ast0} \bar{K}^{\ast0}$  with $1(0^{+})$ and $1(2^{+})$ display attraction, the bound state analysies reveals that the attraction in these channels is too weak to support bound states in single-channel calculations. Consequently, these channels transition into scattering states.

In the context of the diquark-antidiquark configuration, subclusters, each associated with specific color structures, cannot separate directly due to the constraints of the confinement potentials. This implies that channels in this configuration are likely to form resonance states.  Fig.~\ref{diquark-effective} presents the equivalent potential behaviors for various quantum numbers in this configuration. As shown in the figure, the equivalent potential for each single channel exhibits a characteristic behavior: the energy obtained is higher when the two subclusters are either very close or very far apart. Specifically, for systems with quantum numbers $0(1^{+})$, the minimum potential energy occurs at distances of approximately 0.6-0.8 fm, while for $1(0^{+})$, it occurs at 0.7-1 fm.  For systems with $1(1^{+})$ and $1(2^{+})$, the minimum energy is generally found at a separation of around 1 fm. The presence of these minima at intermediate distances suggests the possible formation of resonance states, which are likely to be compact tetraquark states. This provides additional insight into the interaction mechanisms within the diquark-antidiquark configuration and their potential physical significance.

 \begin{figure}
\includegraphics[scale=0.55]{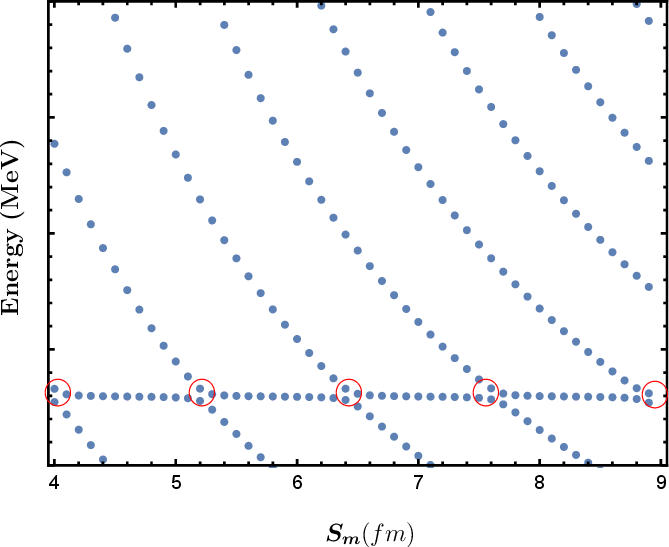}
\caption{A sketch diagram of the resonance shape in the real-scaling method.  }
\label{real_scal1}
\end{figure}

\subsection{Resonance States}
From the aforementioned analysis, it can be inferred that resonance states might exist in the $\ssnn$ tetraquark system. To further explore the possible resonance states in the $\ssnn$ tetraquark system, the real scaling method can be applied to search for and characterize resonances within a multichannel framework that incorporates both meson-meson and diquark-antidiquark configurations~\cite{Kamimura:1981oxj,Kamimura:1977okl}. This method has been successfully employed in electron-atom, electron-molecule, and atom-antiatom systems and is extended to explore genuine resonance states in different systems.   A key aspect of this method is to monitor the variation of estimated energy as the distance $\textbf{S}_{m}$ between two clusters, which serves as a scaling parameter for the basis functions used to represent the system, increases incrementally. With the increased distance between the two clusters, the continuous states, bound states, and resonance states display unique behaviours: continuous states will fall off toward their respective thresholds, while the energies of bound and resonance states will remain stable. Regarding the resonance states, if the scattering state energy is far away from one of the resonance states, the weak coupling will ensure the stability of the resonance state's energy. Conversely, as $\textbf{S}_{m}$ increases, the energy of a scattering state may approach the resonance energy. In this region, the coupling effect between the resonance and scattering state becomes stronger. If $\textbf{S}_{m}$ increases further, the energy gap between these states widens again, and the coupling weakens. This interaction leads to the characteristic "avoided crossing" behaviour, illustrated in Fig.~\ref{real_scal1}, where the energy levels appear to repel each other instead of crossing. Such avoided crossing phenomenon recurs with the increasing of $\textbf{S}_{m}$ due to the continuum nature of the scattering state and the stable energy plateau exhibited by the resonance state before and after the avoided crossing corresponds to the resonance energy.  Additionally, the decay width of the resonance state can be also derived from  
bv\begin{eqnarray}\label{width}
    \Gamma &=& 4|V_{min}(S)|\frac{\sqrt{|k_{r}||k_{c}|} }{|k_{r}-k_{c}|}
\end{eqnarray}
where $k_{r}$ and $k_{c}$ are the slopes of the resonance and scattering states, respectively. Here, $V_{min}(S)$ represents the minimal energy difference between the resonance and the scattering state at the avoided crossing point. This method has been successfully applied to investigate the pentaquark ~\cite{Hiyama:2005cf,Hiyama:2018ukv}, the dibaryon ~\cite{Xia:2021tof}, and the tetraquark systems~\cite{Jin:2020jfc,Liu:2021xje,Liu:2022vyy}. These research findings indirectly validate the rationality and reliability of this method, offering a practical theoretical framework for exploring possible exotic state properties.

\begin{figure}
\includegraphics[scale=0.35]{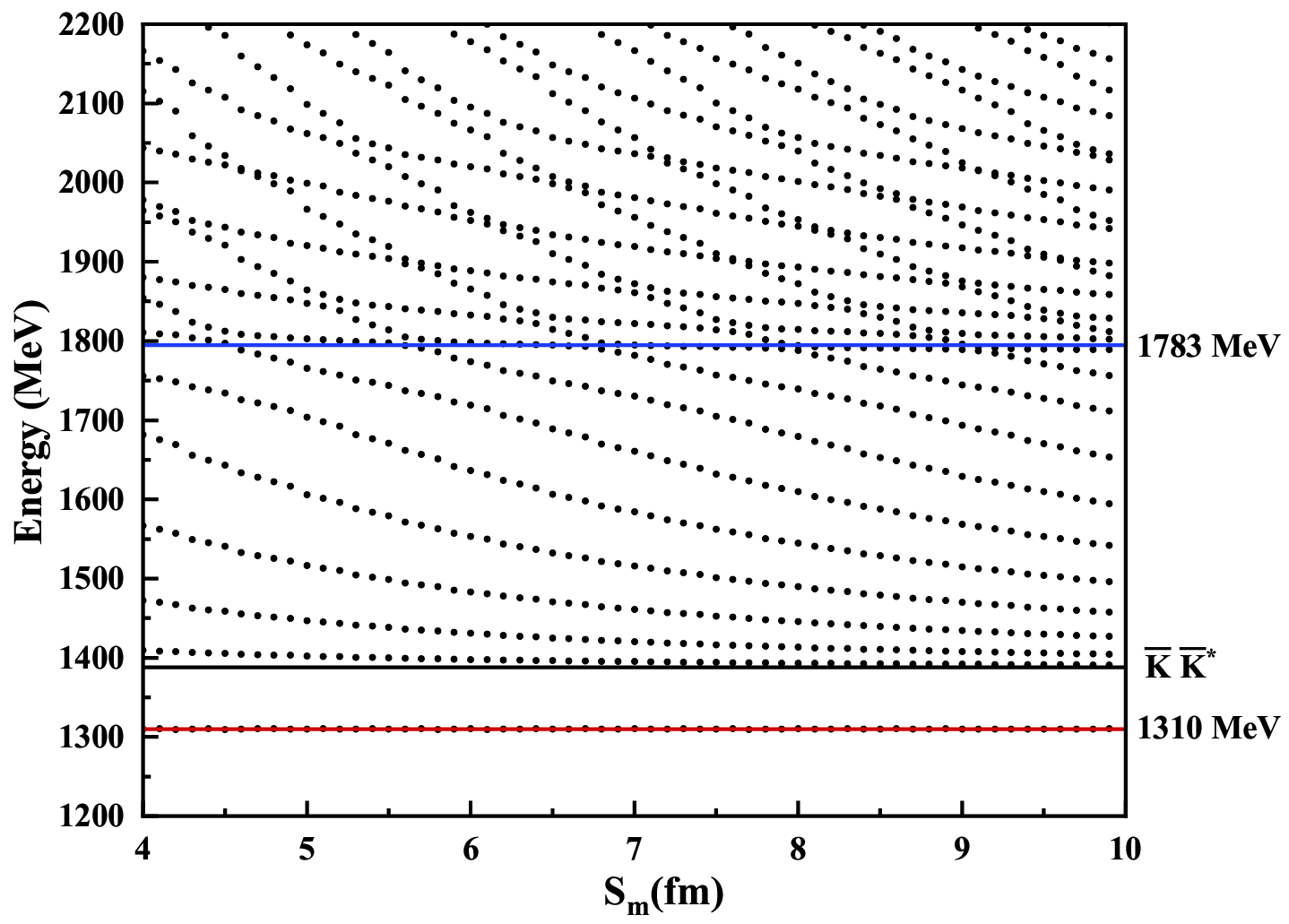}
\caption{The stabilization plots of the energies of the $\ssnn$ tetraquark systems with $I(J^{P})=0(1^{+})$.}
\label{01}
\end{figure}

In the present work, the spatial wave function of the $\ssnn$ tetraquark system with different quantum numbers is expanded using basis functions scaled by the distance $\textbf{S}_{m} (m=1,2,3,\ldots,n)$ between the two constituent clusters. The estimated eigenenergies of the $\ssnn$ tetraquark system with different quantum numbers can be acquired by varying the distance parameter $\textbf{S}_{m}$ from 4 fm to 10 fm, allowing for the identification of possible resonance states through their characteristic behaviour as a function of $\textbf{S}_{m}$.  The relevant results with different quantum numbers in the $\ssnn$ tetraquark system  are presented in Figs~\ref{01}, \ref{10}, \ref{11}, and \ref{12}, respectively. In these figures, black horizontal lines represent the relevant physical thresholds of the involved meson-meson configurations. Red horizontal lines indicate identified bound states, while blue horizontal lines indicate identified genuine resonances. The corresponding estimated masses for these bound and resonance states are marked on the plots.

For the $\ssnn$ tetraquark system with $I(J^{P})=0(1^{+})$,  the channel coupling estimation, including meson-meson and diquark-antidiquark configurations, is performed by gradually increasing the distance parameter $\textbf{S}_{m}$.  As illustrated in Fig.~\ref{01}, with the increasing distance between the two clusters, a bound state and a resonance state progressively emerge. For the identified bound state attributed to the channel coupling, Fig.~\ref{01} indicates that the energy of the bound state remains stable at approximately 1310 MeV. To investigate its properties and structure, we perform detailed calculations of the component proportions for each involved channel and the root-mean-square (RMS) radius of the bound state. The results indicate that the bound state consists of approximately 59\% $\bar{K}^{0}K^{\ast-}$ and $\bar{K}^{\ast0}K^{-}$, 38\% $\bar{K}^{\ast0}K^{\ast-}$, and 3\% diquark-antidiquark configuration. Additionally, its root-mean-square radius is approximately 0.81 fm, suggesting that this bound state is more likely to be a compact tetraquark state. These conclusions regarding the compositions and RMS radius of the bound state are reasonable within the framework of the current QDCSM. The model incorporates characteristics such as quark delocalization and color screening, which generate effective hidden-color channel coupling effects~\cite{Huang:2011kf}, making such results plausible. This implies that the physical channels in the meson-meson configuration not only encompass color singlet structures but also involve hidden color channel structures. Consequently, despite the meson-meson configuration comprising 97\% of the bound state, the small RMS radius suggests a compact structure where hidden-color effects, facilitated by mechanisms like quark delocalization within the QDCSM, play a significant role, even though the overall meson-meson component percentage is high. These considerations make the estimation results reasonable.

\begin{figure}
    \includegraphics[scale=0.35]{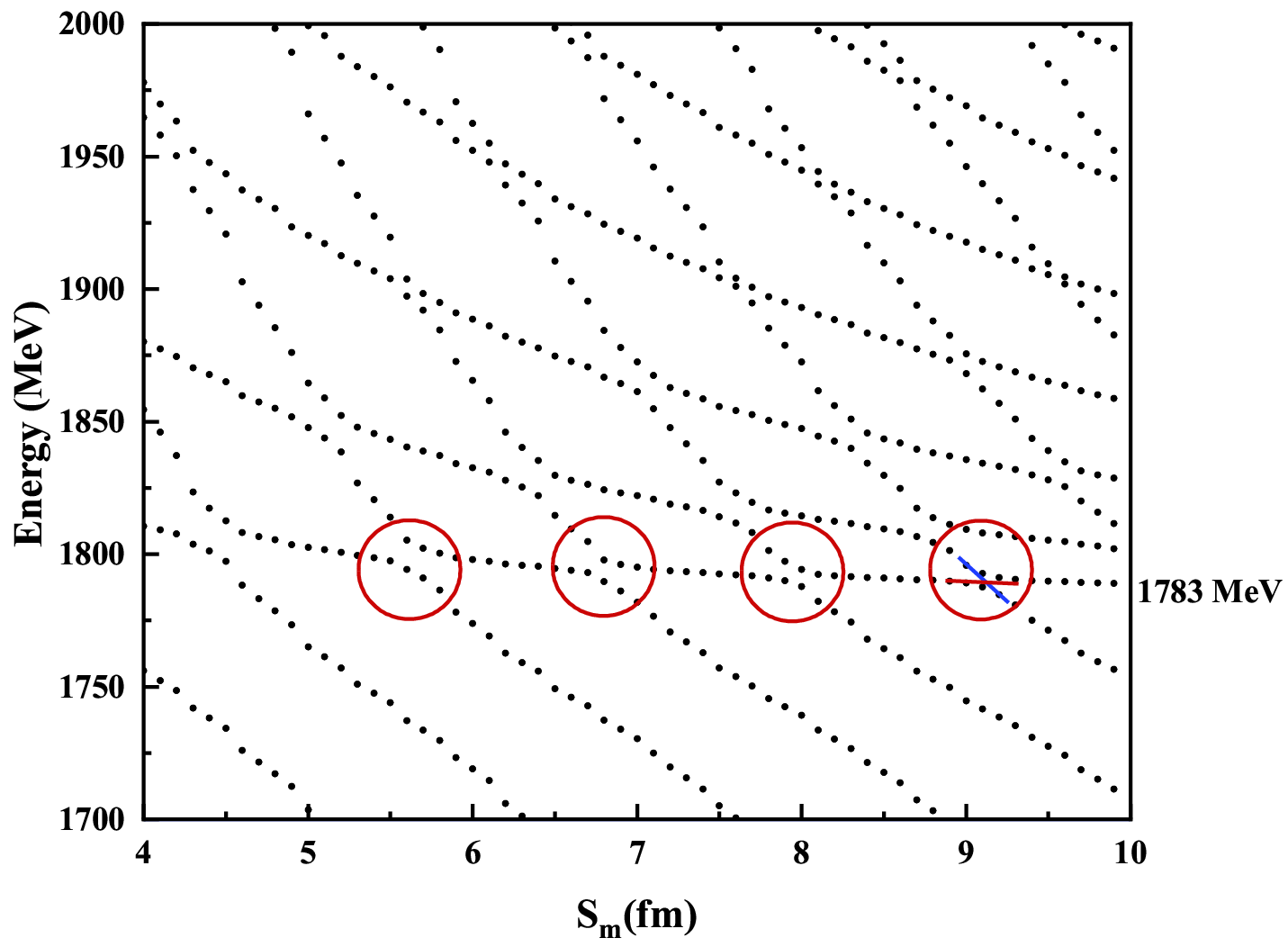}
    \caption{The stabilization plots of the energies of the $\ssnn$ tetraquark systems with $I(J^{P})=0(1^{+})$ in the energy range from 1700 MeV to 2000 MeV. The bule line represents the slope of the scattering states and the red line stands for the solpe of the resonance states. }
    \label{01-range}
\end{figure}

\begin{figure}
    \includegraphics[scale=0.35]{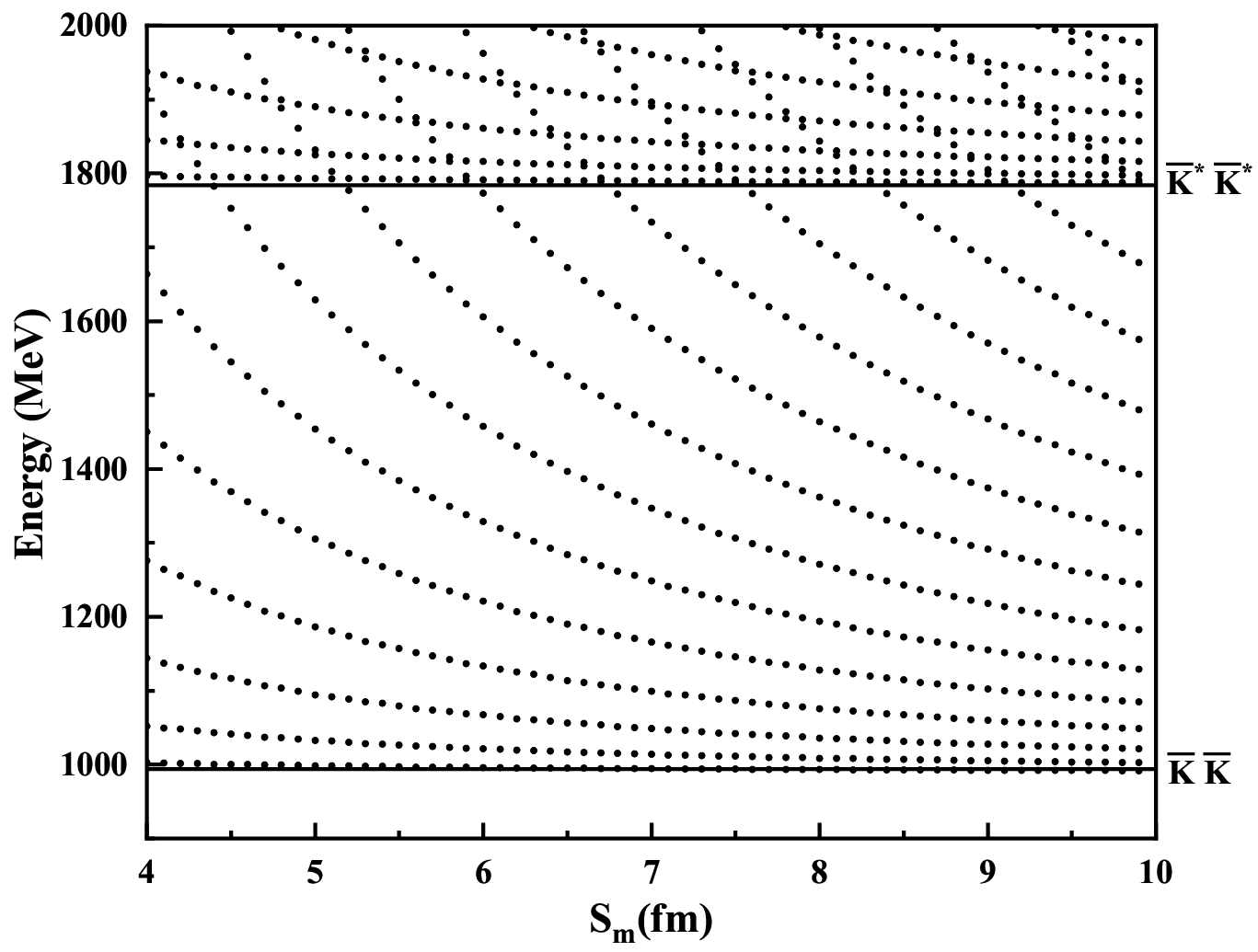}
    \caption{The stabilization plots of the energies of the $\ssnn$ tetraquark systems with $I(J^{P})=1(0^{+})$.}
    \label{10}
\end{figure}

\begin{figure}
    \includegraphics[scale=0.35]{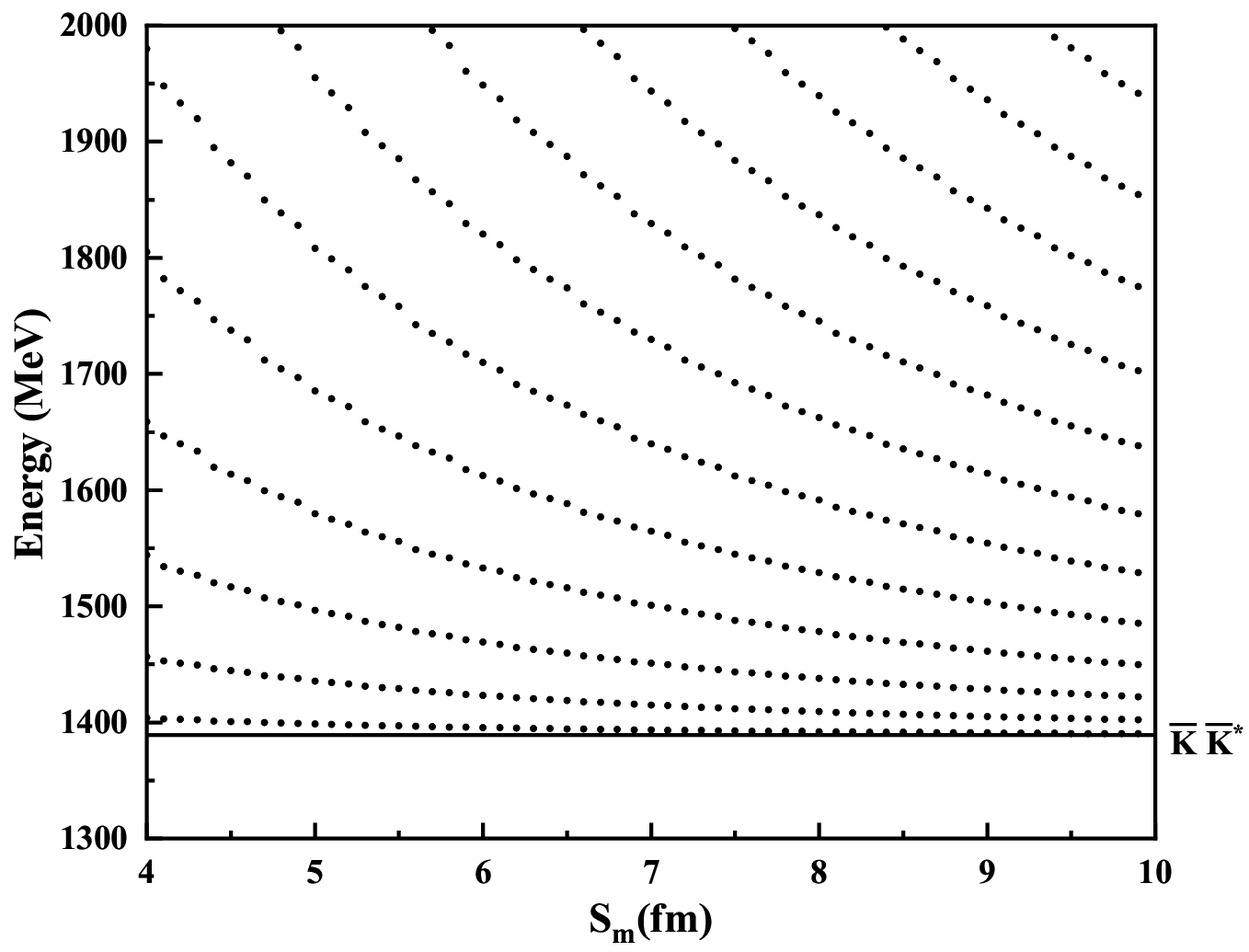}
    \caption{The stabilization plots of the energies of the $\ssnn$ tetraquark systems with $I(J^{P})=1(1^{+})$.}
    \label{11}
\end{figure}

\begin{figure}
    \includegraphics[scale=0.35]{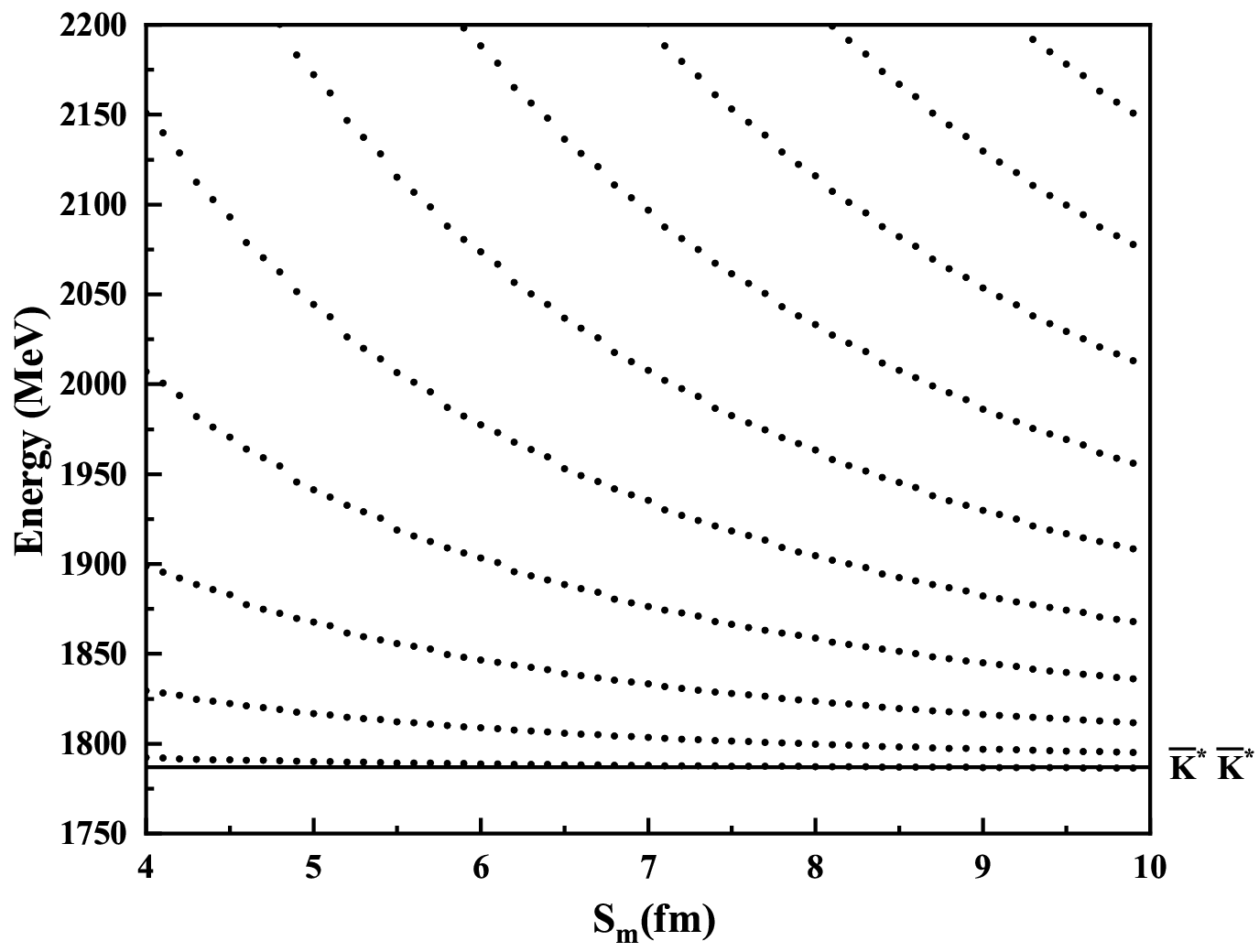}
    \caption{The stabilization plots of the energies of the $\ssnn$ tetraquark systems with $I(J^{P})=1(2^{+})$.}
    \label{12}
\end{figure}

Additionally,  as illustrated in Fig.~\ref{01}, it is evident that the avoided-crossing resonance behavior repeatedly emerges around an energy of approximately 1783 MeV as the finite volume increases,  indicating the possible existence of a resonance state in the $\ssnn$ tetraquark system. To provide a more intuitive visualization of the avoided crossing behavior associated with the resonance state, a magnified view of the energy range from 1700 MeV to 2000 MeV in Fig.~\ref{01-range} is presented.  As depicted in Fig.~\ref{01-range},  the avoided crossing behavior consistently appears near 1783 MeV. Subsequently,  Eq.~\ref{width} is applied to estimate the decay width of the resonance state.  Notably, the estimated width stabilizes with increasing spatial range, indicating that larger spatial ranges produce more reliable results. The estimated decay width is approximately 5.6 MeV.  Moreover, the internal structure of the resonance state is also analyzed. The results indicate that the resonance state with a mass of roughly 1783 MeV is composed of approximately 10\% $\bar{K}^{0} K^{\ast-}$,  10\% $\bar{K}^{\ast0} K^{-}$,  50\% $\bar{K}^{\ast0} K^{\ast-}$, and 30\% diquark-antidiquark configuration. The root mean square radius is approximately 0.58 fm. These observations indicate that the resonance state may be a compact tetraquark state.

For the $\ssnn$ tetraquark system with $I(J^{P})=1(0^+)$, shown in Fig.~\ref{10},  one can find that only two black horizontal lines representing the approximate thresholds for $\bar{K}^{0}\bar{K}^{0}$ and $\bar{K}^{\ast0}\bar{K}^{\ast0}$ are visible, the channel coupling estimations yield no bound states or resonance states. Similar situations are observed for the $\ssnn$ tetraquark system with $I(J^{P})=1(1^+)$ and $I(J^{P})=1(2^+)$. From Figs.~\ref{11} and \ref{12}, a notable feature is that the energy of the continuum state gradually converges toward the respective $\bar{K}^{0}\bar{K}^{\ast0}$ and $\bar{K}^{\ast0}\bar{K}^{\ast0}$ thresholds as the distance parameter $\textbf{S}_{m}$ increases. Again, no bound states or resonance states are found in these channels.

\section{Summary}\label{Sum}
The properties of the  $\ssnn$ tetraquark system with various quantum numbers are systematically investigated using the Quark Delocalization Color Screening Model (QDCSM), with emphasis on effective hadron interactions and the presence of bound or resonance states. Two configurations, meson-meson and diquark-antidiquark, are primarily considered in current studies of the $\ssnn$ tetraquark system.
the single channel estimations, channel coupling within the same configuration, and channel coupling between the two configurations are performed for these configurations. Bound state calculations are conducted using the Resonating Group Method (RGM). Additionally, the stabilization calculation is used to identify potential resonance states in the $\ssnn$ tetraquark system with various quantum numbers.

In the $I(J^{P})=0(1^{+})$ $\ssnn$ tetraquark system, bound states $\bar{K}^{0}K^{\ast-}$, $\bar{K}^{\ast0}K^{-}$, and  $\bar{K}^{\ast0}K^{\ast-}$ with shallow binding energies of approximately -1.5 MeV, -1.5 MeV, and -4.4 MeV, respectively, are revealed by single-channel 
estimations. These bound states, specifically $\bar{K}^{0}K^{\ast-}$ and $\bar{K}^{\ast0}K^{-}$, have been identified in the one-boson-exchange model~\cite{Wang:2024kke} and the chiral quark model ~\cite{Ji:2024znj}, while $\bar{K}^{\ast0}K^{\ast-}$ state is corroborated by the one-boson-exchange model. Furthermore, channel coupling is found to significantly affect the $\ssnn$ tetraquark system. A bound state with an energy of approximately 1310 MeV, consistent with a compact tetraquark configuration, is identified through channel coupling calculations. Additionally, a resonance state with a mass of approximately 1783 MeV and a decay width of about 17 MeV is observed using the stabilization method. For quantum numbers with isospin $I=1$, similar calculations yield no bound or resonance states. Future experiments, such as those conducted by the LHCb and Belle II collaborations through weak decays of B mesons, may probe exotic states in the light quark sector, including the predicted $\bar{K}^{\ast}\bar{K}$ and $\bar{K}^{\ast}\bar{K}^{\ast}$  tetraquark states.

\acknowledgments{This work is supported partly by the National Natural Science Foundation of China under
Contract No. 12175037, No. 12335001, No. 11775118 and No. 11535005. This work is also supported by the Natural Science Foundation for Youths of Henan Province  No. 252300421781.  School-Level Research Projects of Henan Normal unversity (No. 20240304) also supported this work.

\appendix

\section{The wave function of the double-strangeness tetraquark  \label{Sec:App}}

\subsection{The color wave function}
Plenty of color structures in multiquark systems will be available with respect to those of conventional hadrons such as $q\bar{q}$ mesons and $qqq$ baryons. In this section, the goal is to construct the colorless wave function of a 4-quark system.

For the meson-meson configurations, the color wave functions of a $q\bar{q}$ cluster are listed.
\begin{eqnarray}
\nonumber
C^{1}_{[111]} &=& \sqrt{\frac{1}{3}}(r\bar{r}+g\bar{g}+b\bar{b}), \\ \nonumber
C^{2}_{[21]} &=& r\bar{b}, C^{3}_{[21]} =  -r\bar{g},                    \\ \nonumber
C^{4}_{[21]} &=& g\bar{b}, C^{5}_{[21]} =  -b\bar{g},              \\ \nonumber
C^{6}_{[21]} &=& g\bar{r}, C^{7}_{[21]} =   b\bar{r},         \\ \nonumber
C^{8}_{[21]} &=&  \sqrt{\frac{1}{2}}(r\bar{r}-g\bar{g}),       \\
C^{9}_{[21]} &=&  \sqrt{\frac{1}{6}}(-r\bar{r}-g\bar{g}+2b\bar{b}),
\end{eqnarray}

where the subscript [111] and [21] stand for color-singlet ($\textbf{1}_{c}$) and color-octet ($\textbf{8}_{c}$), respectively. So, the $SU(3)_{color}$ wave functions of color-singlet (two color-singlet cluters, $\textbf{1}_{c}\otimes\textbf{1}_{c}$) and hidden-color (two color-octet clusters, $\textbf{8}_{c}\otimes\textbf{8}_{c}$) channels are given, respectively,
\begin{equation}
\chi^{c}_{1} = C^{1}_{[111]}C^{1}_{[111]}, \nonumber
\end{equation}
\begin{equation}
\begin{split}
  \chi^{c}_{2} =&\sqrt{\frac{1}{8}}(C^{2}_{[21]}C^{7}_{[21]}-C^{4}_{[21]}C^{5}_{[21]}-C^{3}_{[21]}C^{6}_{[21]}\\
                &+C^{8}_{[21]}C^{8}_{[21]}-C^{6}_{[21]}C^{3}_{[21]}+C^{9}_{[21]}C^{9}_{[21]}\\
                &-C^{5}_{[21]}C^{4}_{[21]}+C^{7}_{[21]}C^{2}_{[21]}).
\end{split}
\end{equation}

For the diquark-antidiquark structure, the color wave functions of the diquark clusters are given,
\begin{eqnarray}
\nonumber
  C^{1}_{[2]} &=& rr,  C^{2}_{[2]} = \sqrt{\frac{1}{2}}(rg+gr), \\ \nonumber
  C^{3}_{[2]} &=& gg,  C^{4}_{[2]} = \sqrt{\frac{1}{2}}(rb+br),\\ \nonumber
  C^{5}_{[2]} &=& \sqrt{\frac{1}{2}}(gb+bg),  C^{6}_{[2]} = bb, \\ \nonumber
  C^{7}_{[11]} &=& \sqrt{\frac{1}{2}}(rg-gr),  C^{8}_{[11]} = \sqrt{\frac{1}{2}}(rb-br), \\
    C^{9}_{[11]} &=&\sqrt{\frac{1}{2}}(gb-bg).
\end{eqnarray}
While the color wave functions of the antidiquark clusters can be writen as:
\begin{eqnarray}
\nonumber
  C^{1}_{[22]} &=& \bar{r}\bar{r},  C^{2}_{[22]} = -\sqrt{\frac{1}{2}}(\bar{r}\bar{g}+\bar{g}\bar{r}),\\  \nonumber
  C^{3}_{[22]} &=& \bar{g}\bar{g},  C^{4}_{[22]} = \sqrt{\frac{1}{2}}(\bar{r}\bar{b}+\bar{b}\bar{r}), \\  \nonumber
  C^{5}_{[22]} &=& -\sqrt{\frac{1}{2}}(\bar{g}\bar{b}+\bar{b}\bar{g}), C^{6}_{[22]} = \bar{b}\bar{b}, \\ \nonumber
  C^{7}_{[211]}&=& \sqrt{\frac{1}{2}}(\bar{r}\bar{g}-\bar{g}\bar{r}), C^{8}_{[211]} = -\sqrt{\frac{1}{2}}(\bar{r}\bar{b}-\bar{b}\bar{r}), \\
  C^{9}_{[211]} &=& \sqrt{\frac{1}{2}}(\bar{g}\bar{b}-\bar{b}\bar{g}).
\end{eqnarray}
The color-singlet wave functions of the diquark-antidiquark configuration can be the product of color sextet and antisextet clusters ($\textbf{6}_{c}\otimes\bar{\textbf{6}}_{c}$) or the product of color-triplet and antitriplet cluster ($\textbf{3}_{c}\otimes\bar{\textbf{3}}_{c}$), which read,
\begin{equation}
\begin{split}
\chi^{c}_{3} = &\sqrt{\frac{1}{6}}(C^{1}_{[2]}C^{1}_{[22]}-C^{2}_{[2]}C^{[2]}_{[22]}+C^{3}_{[2]}C^{3}_{[22]} \\
               &+C^{4}_{[2]}C^{4}_{[22]}-C^{5}_{[2]}C^{5}_{[22]}+C^{6}_{2}C^{6}_{22}),\nonumber
\end{split}
\end{equation}

\begin{equation}
\begin{split}
  \chi^{c}_{4} =&\sqrt{\frac{1}{3}}(C^{7}_{[11]}C^{7}_{[211]}-C^{8}_{[11]}C^{8}_{[211]}+C^{9}_{[11]}C^{9}_{[211]}).
\end{split}
\end{equation}

\subsection{The flavor wave function}
For the flavor degree of freedom, the different coupling methods generate different flavor wave function. From the Table~\ref{fig1}, the $\ssqq$ tetraquark flavor wave function can be categorized as $F^{i}_{I}$ , the superscript "I" means the total isospin of $\ssqq$ tetraquark states. Distinctive structures are gotten the quark coupling arrange. 

For the $\ssnn$ system,  according to the Fig.~\ref{fig1},  the isospins of the $\ssnn$ system are 0 and 1, respectively.  For the meson-meson structure,  the  coupling orders can be accessed,  which is 
\begin{eqnarray}
     \nonumber F^{1}_{0} &=&  -\sqrt{\frac{1}{2}}(s\bar{d}s\bar{u}-s\bar{u}s\bar{d}) \\
    \nonumber  F^{2}_{1} &=&   (s\bar{d})-(s\bar{d}).
\end{eqnarray}

For the diquark-antidiquark structure, the flavor wave function should be written as
\begin{eqnarray}
 \nonumber    F^{3}_{0}&=& -\sqrt{\frac{1}{2}}(ss\bar{d}\bar{u}-ss\bar{u}\bar{d}) \\
 \nonumber    F^{4}_{1} &=&   (ss)-(\bar{d}\bar{d}).
\end{eqnarray}

%For the $\ssss$ system, the flavor wave function for the meson-meson structure can be obatined by 
%\begin{eqnarray}
%    % \nonumber to remove numbering (before each equation)
%    F^{1}_{0}&=& (s\bar{s})-(s\bar{s})
%\end{eqnarray}
%For the diquark-antidiquark structure, the flavor wave function should be written as
%\begin{eqnarray}
 %   % \nonumber to remove numbering (before each equation)
 %   F^{2}_{0}&=& (ss)-(\bar{s}\bar{s})
%\end{eqnarray}

\subsection{The spin wave function}
For the spin, the total spin $S$ of tetraquark states ranges from 0 to 2. All of them are considered.
The wave functions of two body clusters are
\begin{eqnarray}
\nonumber \chi_{11}&=& \alpha\alpha,\\
\nonumber \chi_{10} &=& \sqrt{\frac{1}{2}}(\alpha\beta+\beta\alpha)\\
\nonumber \chi_{1-1} &=& \beta\beta \\
            \chi_{00} &=& \sqrt{\frac{1}{2}}(\alpha\beta-\beta\alpha)
\end{eqnarray}

Then, the total spin wave functions $S^{i}_{s}$ are obtained by considering the coupling of two subcluster spin wave functions with SU(2) algebra, and the total spin wave functions of four-quark states can be read as
\begin{eqnarray}
\nonumber S^{1}_{0}&=&\chi_{00}\chi_{00}\\
\nonumber S^{2}_{0}&=&\sqrt{\frac{1}{3}}(\chi_{11}\chi_{1-1}-\chi_{10}\chi_{10}+\chi_{1-1}\chi_{11})\\
\nonumber S^{3}_{1}&=&\chi_{00}\chi_{11}\\
\nonumber S^{4}_{1}&=&\chi_{11}\chi_{00}\\
\nonumber S^{5}_{1}&=&\sqrt{\frac{1}{2}}(\chi_{11}\chi_{10}-\chi_{10}\chi_{11}) \\
S^{6}_{2}&=&\chi_{11}\chi_{11}
\end{eqnarray}

\bibliography{SSqq}

\end{document}